\documentclass[aps,prl,reprint,superscriptaddress,floatfix,noeprint]{revtex4-2}

\usepackage{amsmath,amssymb,bm}
\usepackage{graphicx}
\usepackage[colorlinks=true,allcolors=blue]{hyperref}
\usepackage{comment}
\newcommand{\Real}{\operatorname{Re}}
\newcommand{\ee}{\mathbf{1}}
\newcommand{\dd}{\mathrm{d}}
\begin{document}

\title{Diffusion-Induced Stabilization of Nonreciprocal Wealth Dynamics}

\author{Pierre Bousseyroux}
\email{pierre.bousseyroux@polytechnique.edu}
\affiliation{Econophysics Lab, Institut Louis Bachelier, 28 Pl. de la Bourse,
Palais Brongniart, 75002 Paris, France}
\affiliation{LadHyX UMR CNRS 7646, \'{E}cole Polytechnique,
91128 Palaiseau Cedex, France}

\author{Gilles Z\'erah}
\affiliation{Capital Fund Management, 23 Rue de l'Universit\'e,
75007 Paris, France}

\author{Michael Benzaquen}
\affiliation{Econophysics Lab, Institut Louis Bachelier, 28 Pl. de la Bourse,
Palais Brongniart, 75002 Paris, France}
\affiliation{Capital Fund Management, 23 Rue de l'Universit\'e,
75007 Paris, France}

\begin{abstract}
We study the stability of equal wealth shares in a spatial model
combining local sharing, global redistribution, and random
nonreciprocal interactions. On a two-dimensional square lattice
in the infinite-system limit, the effect of local sharing depends
on whether the influence of one site on another tends to have
the same or opposite sign as the reverse influence.
When these influences are nonnegatively correlated, diffusion
reduces the redistribution required for stability, but a positive
rate remains necessary. Negative correlations instead allow
stability without redistribution, with the largest stability
margin at finite diffusion and a return toward marginal stability
at strong diffusion. This nonmonotonicity reflects a competition
between spatial smoothing and the weakening of the stabilizing
contribution of reciprocal anticorrelations. Equal shares can even
remain stable under wealth-concentrating transfers, but only
within an intermediate range of diffusion strengths.
\end{abstract}

\maketitle

Understanding how wealth disparities emerge and persist is a central
question in economic modeling. Beyond differences in individual
returns, wealth dynamics also depend on how economies exchange
resources and influence one another. Stochastic wealth models have shown how random returns and
exchange can generate broad wealth distributions and even
wealth condensation~\cite{BouchaudMezard2000}; see~\cite{YakovenkoRosser2009} for a review. Here, we address a complementary dynamical
question: under what conditions can equal wealth shares remain
stable? In particular, how does local sharing affect this stability,
and how much global redistribution is needed to ensure it?

Local sharing smooths wealth differences between neighboring
economies and therefore tends to favor equality. Its stabilizing
effect is less obvious, however, when economies are also coupled
through heterogeneous, nonreciprocal interactions: the influence
of one economy on another need not match the reverse influence,
either in magnitude or in sign. Random-matrix theory has long
connected the stability of large interacting systems to the
eigenvalues with largest real part \cite{May1972,AllesinaTang2012},
including in economic models \cite{MoranBouchaud2019}. Spatial
structure introduces a further ingredient, as local diffusion
combines with interactions between distant sites. Related
combinations of spatial coupling and random interactions have
been investigated in ecological models
\cite{BaronGalla2020,SalvatoreEtAl2025}.

We study a spatial wealth model combining local sharing, global
redistribution, and random nonlocal interactions. Correlations
between reciprocal interaction coefficients are controlled through
the real Gaussian elliptic ensemble
\cite{Girko1986,SommersEtAl1988}. Linearizing the dynamics around
equal wealth shares yields a lattice Laplacian coupled to a
nonsymmetric random matrix, with redistribution acting as a
uniform spectral shift. Because these operators generally do not
share eigenvectors, their separate spectra do not determine
stability. Recent results on elliptic random perturbations of
self-adjoint matrices
\cite{Ho2022,Zhong2026,BousseyrouxPotters2026}
allow us to determine the right edge of their combined spectrum
and hence the redistribution threshold. 

We find that the effect of diffusion depends qualitatively on
reciprocal correlations. On the two-dimensional square lattice
in the infinite-system limit, nonnegative correlations imply
that stronger diffusion reduces the redistribution required for
stability but never eliminates it. For negative correlations,
equal wealth shares can instead be stable without redistribution,
with the largest stability margin at an intermediate diffusion
strength. Beyond this optimum, stronger diffusion brings the
system back toward marginal stability. This nonmonotonic behavior
reflects a competition between spatial smoothing and the weakening
of the stabilizing contribution of reciprocal anticorrelations.
Equal wealth shares can even remain stable in the presence of
wealth-concentrating transfers, but then only within an
intermediate range of diffusion strengths.

\paragraph{Model.}
Consider $N=L^2$ local economies on a periodic square lattice, each with initially positive wealth $W_i$. At each site, a fraction $x\in[0,1)$ of wealth is committed to local projects, whose turnover time sets the unit of time. These projects consume wealth at rate $xW_i$ and generate a gross return at rate $rxW_i$, with $r\geq0$. The return is shared equally among the originating site and its four nearest neighbors.
The remaining wealth, \((1-x)W_i\), is exposed to interactions beyond
local sharing, described by coefficients \(J_{ij}\). 

Specifically, we assume that wealth evolves as
\begin{equation}
\begin{split}
\dot W_i={}&-xW_i+\frac{rx}{5}
\left(W_i+\sum_{j\sim i}W_j\right)\\
&+(1-x)\sum_j J_{ij}(\bar W-W_j),
\end{split}
\label{eq:map}
\end{equation}
where \(j\sim i\) denotes the four nearest neighbors of site \(i\)
and \(\bar W=N^{-1}\sum_i W_i\) is the mean wealth.
Defining the lattice Laplacian by
\((\Delta W)_i=\sum_{j\sim i}(W_j-W_i)\)
and writing \(J_{ij}=\sigma X_{ij}+\gamma\delta_{ij}\),
with \(\sigma>0\) and \(\gamma\in\mathbb R\),
we rewrite Eq.~\eqref{eq:map} as
\begin{eqnarray}
\dot W_i&=& x(r-1)W_i+\frac{rx}{5}(\Delta W)_i
+\gamma(1-x)(\bar W-W_i)\nonumber \\
&& + \, \sigma(1-x)\sum_j X_{ij}(\bar W-W_j).
\label{eq:model}
\end{eqnarray}
The first term in~\eqref{eq:model} makes wealth at every site grow or decline at the same
proportional rate \(x(r-1)\), leaving relative wealth shares unchanged. The second describes wealth diffusion through local sharing. The third term redistributes wealth from richer (above-average) to poorer (below-average) sites when
\(\gamma>0\) and amplifies wealth differences when \(\gamma<0\). The last term describes nonlocal interactions through which wealth
deviations at other sites affect wealth growth at site \(i\).
The coefficients \(X_{ij}\) encode the sign and relative strength
of these effects, while \(\sigma\) sets their overall scale. 

We draw the random matrix \(X=X_\tau\) from the real Gaussian elliptic
ensemble and hold its realization fixed in time~\cite{Girko1986,SommersEtAl1988}.
Its entries have zero mean, and the off-diagonal variance is \(1/N\),
which keeps the spectrum of order one as \(N\) grows. The diagonal
entries are of order \(N^{-1/2}\) and do not affect the large-\(N\)
results [Supplemental Material (SM), Sec.~\ref{sm:projection}].
The only correlations are between reciprocal coefficients,
\(\operatorname{Corr}(X_{ij},X_{ji})=\tau\), with
\(-1\leq\tau\leq1\). This parameter interpolates between symmetric
interactions at \(\tau=1\), where reciprocal effects have equal
magnitude and the same sign, and antisymmetric interactions at
\(\tau=-1\), where they have equal magnitude and opposite signs,
as in predator--prey interactions. 

When all sites have the same wealth, the diffusion, redistribution,
and nonlocal interaction terms in Eq.~\eqref{eq:model} vanish.
Wealth then evolves uniformly as
\(W_i(t)=w(0)e^{x(r-1)t}\), with \(w(0)>0\).

\paragraph{Wealth shares and linear stability.}

To distinguish changes in wealth inequality from overall growth,
we introduce the wealth shares
\begin{equation}
p_i=\frac{W_i}{\sum_j W_j}.
\end{equation}
The uniform trajectory becomes a stationary equal-share state,
\(p_i=1/N\). Writing \(p_i=1/N+u_i\), with \(\sum_i u_i=0\),
and linearizing around this state gives
\begin{equation}
\dot u=\mathcal L u,
\qquad
\mathcal L=(1-x)\sigma
\left[\chi\Delta+\widehat X_\tau-\beta I\right],
\label{eq:jacobian}
\end{equation}
where
\begin{equation}
\chi=\frac{rx}{5(1-x)\sigma},
\qquad
\beta=\frac{\gamma}{\sigma}.
\label{eq:reduced}
\end{equation}
The dimensionless parameters \(\chi\) and \(\beta\) measure the
strengths of diffusion and redistribution relative to that of
nonlocal interactions, with \(\beta<0\) corresponding to wealth
concentration.

All operators in Eq.~\eqref{eq:jacobian} act on the zero-sum subspace
\(\ee^\perp\), where \(\ee=(1,\ldots,1)^\top\), and \(I\) denotes
the identity on this subspace. 
The effective interaction matrix \(\widehat X_\tau\) is obtained
by projecting \(-X_\tau\) onto \(\ee^\perp\): this projection
subtracts the mean component of the interaction-induced changes,
ensuring that the share perturbations continue to sum to zero.
The minus sign comes from \(\bar W-W_j\) in Eq.~\eqref{eq:map}.
In an orthonormal basis of \(\ee^\perp\), \(\widehat X_\tau\)
remains a Gaussian elliptic matrix of dimension \(N-1\), with
the same reciprocal correlation \(\tau\) and off-diagonal
variance \(1/N\) (SM, Sec.~\ref{sm:projection}).
On the periodic square lattice, the eigenmodes of \(\Delta\)
are Fourier modes with wavevectors
\(k=\frac{2\pi}{L}(a,b)\), where \(a,b=0,\ldots,L-1\).
Their eigenvalues are \(-\omega(k)\), with
\[
\omega(k)=4-2\cos k_x-2\cos k_y.
\]
The uniform mode \(k=0\) is excluded from \(\ee^\perp\).

For a finite realization, the equal-share state is linearly stable
if all eigenvalues of \(\mathcal L\) have strictly negative real parts. Because the nonlocal interactions can have either sign, initially
positive wealth need not remain positive for arbitrary initial
conditions. Near a linearly stable equal-share state, however,
wealth remains positive for all times if the initial perturbation
from a positive uniform state is sufficiently small. In the limit \(L\to\infty\), let \(\mathcal S_{\chi,\tau}\)
denote the limiting spectral support of
\(\chi\Delta+\widehat X_\tau\), and define its right edge by
\begin{equation}
\Phi(\chi,\tau)
=\sup\{\Real z:z\in\mathcal S_{\chi,\tau}\}.
\label{eq:phi-def}
\end{equation}
The right edge of the limiting spectrum of \(\mathcal L\) is then
\begin{equation}
\begin{split}
\Lambda
&=(1-x)\sigma\bigl[\Phi(\chi,\tau)-\beta\bigr]\\
&=(1-x)\bigl[\sigma\Phi(\chi,\tau)-\gamma\bigr].
\end{split}
\label{eq:Lambda}
\end{equation}
Thus, the limiting spectral support lies strictly in the left
half-plane if and only if
\begin{equation}
\Lambda<0
\quad\Longleftrightarrow\quad
\beta>\Phi(\chi,\tau).
\end{equation}

\begin{figure*}[t]
\centering
\includegraphics[width=\textwidth]{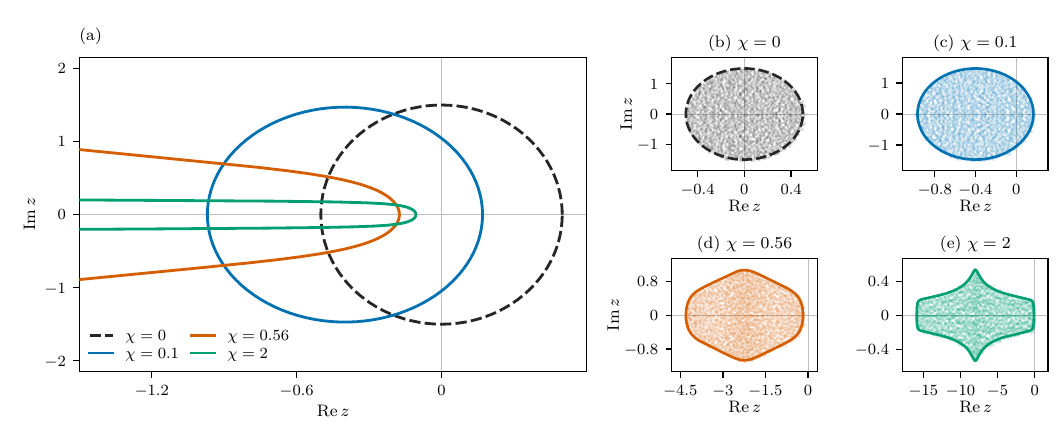}
\caption{\label{fig:spectrum}
(a) Boundary of the limiting spectrum of
\(\chi\Delta+\widehat X_\tau\) at \(\tau=-0.5\), for
\(\chi=0\), \(0.1\), \(0.56\), and \(2\), computed from
Eqs.~\eqref{eq:v-boundary} and \eqref{eq:full-boundary}.
The horizontal axis starts at \(-1.5\).
The dashed curve is the boundary of the elliptic law at
\(\chi=0\), Eq.~\eqref{eq:ellipse}.
(b)--(e) Eigenvalues on \(\ee^\perp\) for the same realization
of \(\widehat X_\tau\), with \(L=50\) (\(N=2500\)) and
\(\tau=-0.5\), at the four diffusion strengths shown in (a),
respectively. The value \(\chi\simeq0.56\) minimizes the right
spectral edge at \(\tau=-0.5\).
The curve in each panel is the corresponding limiting boundary
from~(a).}
\end{figure*}

\paragraph{Spectral boundary.}
Following Ref.~\cite{BousseyrouxPotters2026}, we parameterize the
boundary of \(\mathcal S_{\chi,\tau}\) by a real variable \(s\)
(SM, Sec.~\ref{sm:boundary}). For each admissible \(s\), we define
\(v=v(s)\geq0\) as the solution of
\begin{equation}
\int_{[-\pi,\pi]^2}\frac{\dd^2k}{(2\pi)^2}
\frac{1}{[s+\chi\omega(k)]^2+v^2}=1.
\label{eq:v-boundary}
\end{equation}
The corresponding boundary points are
\begin{equation}
\begin{split}
z_\pm(s)={}&s+\tau\int_{[-\pi,\pi]^2}\frac{\dd^2k}{(2\pi)^2}
\frac{s+\chi\omega(k)}{[s+\chi\omega(k)]^2+v^2}\\
&\pm i(1-\tau)v,
\end{split}
\label{eq:full-boundary}
\end{equation}
where the two signs describe complex-conjugate branches.
Without diffusion (\(\chi=0\)), these equations recover the
boundary of the elliptic law, with real and imaginary semiaxes
\(1+\tau\) and \(1-\tau\), respectively:
\begin{equation}
\frac{(\Real z)^2}{(1+\tau)^2}
+\frac{(\operatorname{Im}z)^2}{(1-\tau)^2}=1
\qquad (|\tau|<1).
\label{eq:ellipse}
\end{equation}
At \(\tau=1\), the support lies on the real axis for all \(\chi\).
At \(\tau=-1\), it lies on the imaginary axis only in the absence
of diffusion.
Figure~\ref{fig:spectrum}(a) illustrates how diffusion deforms the
boundary at \(\tau=-0.5\), while
Figs.~\ref{fig:spectrum}(b)--(e) compare each limiting boundary
with the eigenvalues of a single realization.

\paragraph{Right edge.}
For \(|\tau|<1\), \(\Real z\) increases with \(s\) along the
boundary, so the right edge corresponds to the largest admissible
value of \(s\), denoted by \(\xi\). On the square lattice, this
endpoint satisfies \(v=0\) (SM, Sec.~\ref{sm:boundary}).
Writing \(\langle f(\omega)\rangle\) for the average of
\(f(\omega(k))\) over wavevectors, we obtain
\begin{equation}
\begin{gathered}
\langle(\xi+\chi\omega)^{-2}\rangle=1,\qquad \xi>0,\\
\Phi=\xi+\tau G,\qquad
G=\langle(\xi+\chi\omega)^{-1}\rangle.
\end{gathered}
\label{eq:right-general}
\end{equation}
Since neither \(\xi\) nor \(G\) depends on \(\tau\), the right
edge varies linearly with \(\tau\), with
\(\partial_\tau\Phi=G\) at fixed \(\chi\).
Both \(\xi\) and \(G\) decrease with diffusion:
\(\xi'(\chi)<0\) and \(G'(\chi)<0\).
These inequalities hold for any nonconstant dispersion wherever
a positive solution \(\xi\) exists
(SM, Sec.~\ref{sm:monotonicity}).

On the square lattice, Eq.~\eqref{eq:right-general} admits a
unique positive solution \(\xi\) for every \(\chi>0\), and the
lattice Green function~\cite{Guttmann2010} gives
(SM, Sec.~\ref{sm:green})
\begin{equation}
\begin{gathered}
\frac{2E(m)}{\pi\xi(\xi+8\chi)}=1,\qquad
m=\left(\frac{4\chi}{\xi+4\chi}\right)^2,\\
G=\frac{2K(m)}{\pi(\xi+4\chi)},
\end{gathered}
\label{eq:KE}
\end{equation}
where \(K(m)\) and \(E(m)\) are the complete elliptic integrals
of the first and second kinds, respectively:

\begin{eqnarray}
    K(m)&=&\int_0^{\pi/2}\frac{\dd\theta}{\sqrt{1-m\sin^2\theta}},\nonumber \\
E(m)&=&\int_0^{\pi/2}\sqrt{1-m\sin^2\theta}\,\dd\theta.
\end{eqnarray}
At small \(\chi\),
\begin{eqnarray}
\xi=1-4\chi+O(\chi^2),\qquad G=1+O(\chi^2),
\end{eqnarray}
so \(\Phi(0,\tau)=1+\tau\) and the initial slope is \(-4\)
for every \(\tau\).

\paragraph{Two-dimensional strong-diffusion limit.}
Near \(k=0\), the square-lattice dispersion satisfies
\(\omega(k)=|k|^2+O(|k|^4)\). Using this long-wavelength
expansion together with the lattice Green function, we obtain
(SM, Sec.~\ref{sm:green})
\begin{equation}
\begin{gathered}
\xi\underset{\chi\to\infty}{\sim}\frac{1}{4\pi\chi},\\
G(\chi)=\frac{\log\chi+\log(8\sqrt{2\pi})}{2\pi\chi}
+o(\chi^{-1}).
\end{gathered}
\label{eq:strong}
\end{equation}
Thus, for fixed \(\tau\), the right edge \(\Phi\) approaches
zero from below when \(\tau<0\) and from above when \(\tau\geq0\).
Resolving the long-wavelength modes that control this behavior
requires \(L\gg4\pi^{3/2}\chi\)
(SM, Sec.~\ref{sm:finite-size}).
The order of limits is therefore important: at fixed finite \(L\),
the largest real part of the eigenvalues of
\(\chi\Delta+\widehat X_\tau\) on \(\ee^\perp\) tends to
\(-\infty\) as \(\chi\to\infty\), whereas taking \(L\to\infty\)
first yields a spectral edge that tends to zero.

\begin{figure}[!t]
\centering
\includegraphics[width=\columnwidth]{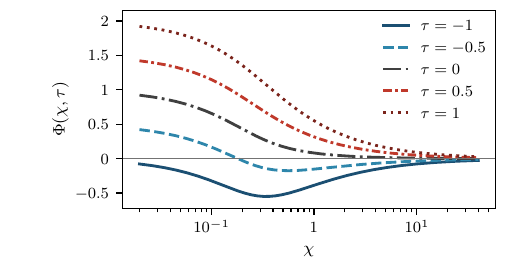}
\caption{\label{fig:cuts}
Right edge \(\Phi(\chi,\tau)\) of the limiting spectrum,
Eq.~\eqref{eq:phi-def}, as a function of the relative diffusion
strength \(\chi\), Eq.~\eqref{eq:reduced}, for the values of
\(\tau\) indicated in the legend. Curves are computed from
Eqs.~\eqref{eq:right-general} and \eqref{eq:KE}.
For \(\tau\geq0\), the edge decreases monotonically and remains
positive. For \(\tau<0\), it reaches a global minimum at finite
\(\chi\) and approaches zero from below as \(\chi\to\infty\).
At fixed \(\chi\), \(\partial_\tau\Phi=G(\chi)\).}
\end{figure}

\paragraph{Finite-diffusion minimum.}
Since \(\partial_\chi\Phi=\xi'(\chi)+\tau G'(\chi)\) and both
derivatives are negative, the edge decreases monotonically for
\(\tau\geq0\). For \(\tau<0\), by contrast, the edge initially
decreases with slope \(-4\), becomes negative, and approaches
zero from below as \(\chi\to\infty\). It therefore attains a
global minimum at a finite, positive diffusion strength;
we denote one such minimizer by \(\chi_\star(\tau)\).
For \(-1<\tau<0\), the edge crosses zero at a unique diffusion
strength \(\chi_0(\tau)\), with \(\Phi<0\) for
\(\chi>\chi_0(\tau)\); we set \(\chi_0(-1)=0\)
(SM, Sec.~\ref{sm:green}).
Figure~\ref{fig:cuts} illustrates these two regimes.

As the negative reciprocal correlation weakens, the optimal
diffusion strength grows exponentially
(SM, Sec.~\ref{sm:finite-size}):
\[
\chi_\star(\tau)
\underset{\tau\to0^-}{\sim}
\frac{e}{8\sqrt{2\pi}}\exp\left(\frac{1}{2|\tau|}\right).
\]

\paragraph{Redistribution.}
Redistribution shifts every eigenvalue of the dimensionless
stability operator by \(-\beta\). The infimum of nonnegative
redistribution rates ensuring strict stability is
\begin{equation}
\gamma_c=\sigma\max\{0,\Phi(\chi,\tau)\}.
\label{eq:redistribution-level}
\end{equation}
Strict stability requires \(\gamma>\sigma\Phi\).
When \(\Phi\geq0\), the threshold \(\gamma=\gamma_c\) is
marginal. When \(\Phi<0\), equal shares are stable without
redistribution and even under wealth concentration, provided
\(\sigma\Phi<\gamma<0\).
In the region \(\Phi>0\),
\begin{equation}
\partial_\tau\gamma_c=\sigma G(\chi)>0,\qquad
\partial_\chi\partial_\tau\gamma_c=\sigma G'(\chi)<0.
\end{equation}
Thus, increasing reciprocal correlation raises the redistribution
threshold, while diffusion reduces its sensitivity to that correlation.

\paragraph{Discussion.}
Let us summarize the main findings. Local sharing can stabilize
equal wealth shares, but its effect depends on the correlations
between reciprocal nonlocal interactions. When these correlations
are nonnegative, stronger sharing progressively reduces the
redistribution needed for stability, but never eliminates it
in the infinite-system limit. When reciprocal effects tend to
have opposite signs (\(\tau<0\)), equal shares can instead be
stable without redistribution. In this case, stability is
strongest at an intermediate sharing strength: increasing
diffusion beyond this optimum reduces the stability margin,
although the state remains stable at every larger finite
diffusion strength.

This nonmonotonic behavior reflects two competing effects.
The decomposition \(\Phi=\xi+\tau G\),
Eq.~\eqref{eq:right-general}, separates a contribution
\(\xi\) that decreases with diffusion from a contribution
\(\tau G\) associated with reciprocal correlations.
Recall that a smaller \(\Phi\) means a larger stability
margin at fixed redistribution. Diffusion lowers \(\xi\),
which favors stability, but also lowers \(G\), reducing
the influence of reciprocal correlations. For \(\tau<0\),
these correlations provide a stabilizing contribution,
which diffusion progressively weakens. The competition
between these effects produces the finite-diffusion minimum.
For \(\tau\geq0\), both effects instead lower the spectral
edge, so stronger diffusion always helps.

\begin{figure}[!t]
\centering
\includegraphics[width=\columnwidth]{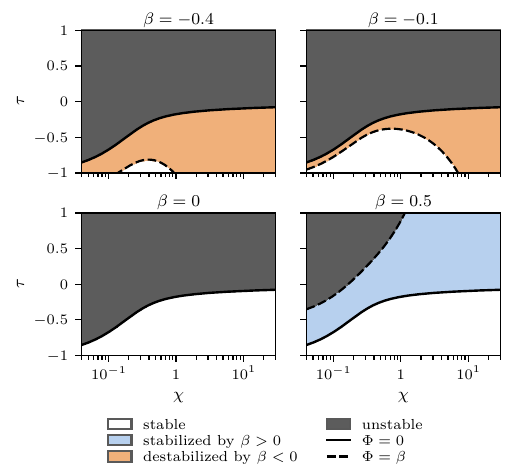}
\caption{\label{fig:sign}
Linear stability of equal wealth shares in the
\((\chi,\tau)\) plane for \(\beta=-0.4,-0.1,0,\) and \(0.5\)
(top left, top right, bottom left, and bottom right, respectively),
where \(\beta=\gamma/\sigma\).
The solid curve marks the stability threshold without
redistribution (\(\Phi=0\)); the dashed curve marks the
threshold at the indicated \(\beta\) (\(\Phi=\beta\)).
The curves coincide when \(\beta=0\).
Colors compare stability at the indicated \(\beta\) with
the reference case of no redistribution:
white, stable in both cases;
dark gray, unstable in both cases;
light blue, unstable without redistribution but stabilized
by positive redistribution (\(\beta>0\));
light orange, stable without redistribution but destabilized
by wealth concentration (\(\beta<0\)).}
\end{figure}

The consequences for redistribution are illustrated in
Fig.~\ref{fig:sign}. Equal shares are stable when
\(\Phi<\beta\), so positive redistribution expands the stable
region beyond the region \(\Phi<0\) where local sharing and
nonlocal interactions suffice on their own.
Conversely, negative redistribution amplifies wealth
differences. Nevertheless, when negative reciprocal
correlations are strong enough, equal shares can remain
linearly stable even under this concentrating tendency.
For a fixed negative redistribution rate, this stability
is restricted to intermediate diffusion strengths:
too little sharing fails to stabilize the system,
while too much weakens the stabilizing contribution
of reciprocal correlations.

These conclusions have two important qualifications.
First, the strong-diffusion behavior depends on spatial
geometry, through the long-wavelength modes of the
Laplacian. Although diffusion reduces sensitivity to
reciprocal correlations more generally, the return of
the spectral edge toward zero from below occurs for
every \(\tau<0\) on the square lattice, but only for
\(\tau<-1/2\) on the one-dimensional cycle
(SM, Sec.~\ref{sm:cycle}).

Second, the finite-diffusion optimum is an infinite-system
result that can require very large lattices to observe.
As negative reciprocal correlations weaken,
\(\chi_\star\) grows exponentially, and resolving the
predicted minimum requires
\(L\gg4\pi^{3/2}\chi_\star\).
The required lattice size therefore grows exponentially
as well (SM, Sec.~\ref{sm:finite-size}).
At fixed finite size, sufficiently strong diffusion
ultimately restores increasingly strong stability,
rather than returning the system toward marginality.

\paragraph{Acknowledgments}
This research was conducted within the Econophysics \& Complex Systems
Research Chair, under the aegis of the Fondation du Risque and Capital Fund
Management. {The authors acknowledge the use of multiple generative AI models, including Claude Opus and Claude Sonnet (Anthropic), and GPT Sol (OpenAI), to assist with research and manuscript preparation. All results and
interpretations were checked by the authors, who take full
responsibility for the content. }

\bibliographystyle{apsrev4-2}
\bibliography{references}

\clearpage
\onecolumngrid
\phantomsection
\pdfbookmark[0]{Supplemental Material}{supplemental-material}

\setcounter{section}{0}
\setcounter{subsection}{0}
\setcounter{equation}{0}
\renewcommand{\thesection}{\Roman{section}}
\renewcommand{\theequation}{S\arabic{equation}}
\providecommand{\theHsection}{}
\renewcommand{\theHsection}{SM.\arabic{section}}
\providecommand{\theHequation}{}
\renewcommand{\theHequation}{SM.\arabic{equation}}

\begin{center}
\textbf{Supplemental Material for}\\[0.5em]
{``Diffusion-Induced Stabilization of Nonreciprocal Wealth Dynamics''}\\[0.5em]
Pierre Bousseyroux, Gilles Z\'erah, and Michael Benzaquen
\end{center}

\section{Projection onto wealth-share perturbations}\label{sm:projection}

Let $\ee=(1,\ldots,1)^\top$ and
$\Pi=I-\ee\ee^\top/N$.
Equation~\eqref{eq:map} can be written as
$\dot W=\mathcal K W$, with
\[
\mathcal K=x(r-1)I+\frac{rx}{5}\Delta
-(1-x)(\sigma X_\tau+\gamma I)\Pi.
\]

For $p=W/(\ee^\top W)$, the exact share dynamics is
$\dot p=\mathcal F(p)$, where
$\mathcal F(p)=\mathcal Kp-p(\ee^\top\mathcal Kp)$.
Since $\mathcal K\ee=x(r-1)\ee$, the equal-share
state $q=\ee/N$ is stationary. Its Jacobian on
$\ee^\perp$ is
\[
D\mathcal F_q
=\left.\Pi[\mathcal K-x(r-1)I]
\right|_{\ee^\perp}
=\frac{rx}{5}\Delta
+(1-x)(\sigma\widehat X_\tau-\gamma I),
\]
where
\[
\widehat X_\tau
=-(\Pi X_\tau\Pi)|_{\ee^\perp}.
\]
The left projection arises from share
normalization, while the right projection
centers the wealth deviations.

Let $U\in\mathbb R^{N\times(N-1)}$ have
orthonormal columns spanning $\ee^\perp$.
In this basis, $\widehat X_\tau$ is represented
by $Y=-U^\top X_\tau U$.
Using the Gaussian covariance
\[
\mathbb E[(X_\tau)_{ij}(X_\tau)_{kl}]
=\frac{\delta_{ik}\delta_{jl}
+\tau\delta_{il}\delta_{jk}}{N},
\]
orthonormality gives
\[
\mathbb E[Y_{\alpha\beta}Y_{\mu\nu}]
=\frac{\delta_{\alpha\mu}\delta_{\beta\nu}
+\tau\delta_{\alpha\nu}\delta_{\beta\mu}}{N}.
\]
Thus $Y$ is exactly a Gaussian elliptic
matrix of dimension $N-1$, with the same
reciprocal correlation $\tau$.
Its normalization differs from the standard
$1/(N-1)$ convention only by the matrix
factor $\sqrt{(N-1)/N}$.

\section{Spectral boundary and right edge}
\label{sm:boundary}

We apply the spectral-boundary equations of
Ref.~\cite{BousseyrouxPotters2026} to
$\chi\Delta+\widehat X_\tau$.
Since the eigenvalues of $\chi\Delta$ are
$-\chi\omega(k)$, these equations give
\[
\left\langle
\frac{1}{|w+\chi\omega|^2}
\right\rangle=1,
\qquad
z=w+\tau\left\langle
\frac{1}{w+\chi\omega}
\right\rangle,
\]
where $w\in\mathbb C$ is an auxiliary parameter
and the brackets denote the average over
lattice wavevectors.

Writing $w=s+iv$ gives
Eqs.~\eqref{eq:v-boundary}
and \eqref{eq:full-boundary} of the Letter.

\paragraph{Rightmost point of the boundary.}

We first show that the real part of the spectral
boundary increases with $s$.

For $v>0$, let
\[
a=s+\chi\omega,\qquad D=a^2+v^2,
\]
and introduce
\[
B=\langle D^{-2}\rangle,\quad
C=\langle aD^{-2}\rangle,\quad
A_2=\langle a^2D^{-2}\rangle.
\]

Differentiating $\langle D^{-1}\rangle=1$
along the boundary gives $vv'=-C/B$.
Consequently, for $F=\langle a/D\rangle$,
\[
F'=1-2\left(A_2-\frac{C^2}{B}\right).
\]

By Cauchy--Schwarz and $a^2\leq D$,
\[
0\leq A_2-\frac{C^2}{B}\leq1.
\]
It follows that
\[
\frac{\dd\,\Real z}{\dd s}
=1+\tau F'\geq1-|\tau|>0
\qquad (|\tau|<1).
\]

The rightmost point therefore corresponds to
the largest admissible value of $s$.

For $s>0$, the quantity
\[
\left\langle(s+\chi\omega)^{-2}\right\rangle
\]
decreases continuously from infinity to zero
as $s$ increases. There is thus a unique
$\xi>0$ satisfying
\[
\left\langle
(\xi+\chi\omega)^{-2}
\right\rangle=1.
\]

No boundary point exists at larger $s$,
and the endpoint is reached at $v=0$.
The right spectral edge is therefore
\[
\Phi(\chi,\tau)=\xi+\tau G,
\qquad
G=\left\langle
(\xi+\chi\omega)^{-1}
\right\rangle,
\]
as stated in Eq.~\eqref{eq:right-general}.

\section{Monotonicity of \texorpdfstring{\(\xi\)}{xi} and \texorpdfstring{\(G\)}{G}}\label{sm:monotonicity}

We show that both $\xi$ and $G$ decrease
with diffusion. Introducing
\[
R=\xi(\chi)+\chi\omega,
\qquad
M_n=\langle R^{-n}\rangle,
\]
the defining equations read $M_2=1$
and $G=M_1$.

Differentiating with respect to $\chi$ gives
\[
\xi'=\frac{\xi-M_3^{-1}}{\chi},
\qquad
G'=\frac{M_3^{-1}-G}{\chi}.
\]
Since $R\geq\xi$, with strict inequality
except at the zero modes,
\[
\xi M_3<M_2=1,
\]
so that $\xi'<0$.

Furthermore, Cauchy--Schwarz gives
\[
GM_3>M_2^2=1,
\]
which implies $G'<0$.

These results do not rely on the specific
square-lattice dispersion: they hold for
any nonconstant, nonnegative dispersion
for which the positive solution exists.

\section{Square-lattice Green function and strong-diffusion asymptotics}\label{sm:green}

With \(\mathbb T^2=[-\pi,\pi]^2\), write
\[
M_n(\xi,\chi)=\int_{\mathbb T^2}\frac{\dd^2k}{(2\pi)^2}
[\xi+\chi\omega(k)]^{-n},
\]
with \(\omega(k)=4-2\cos k_x-2\cos k_y\). With \(d=\xi+4\chi\), the
square-lattice Green function~\cite{Guttmann2010} is
\[
M_1(\xi,\chi)=\frac{2K(m)}{\pi d},
\qquad
m=\left(\frac{4\chi}{d}\right)^2,
\]
where \(K\) and \(E\) are the complete elliptic integrals in the parameter
convention of the Letter. At fixed \(\chi\),
\[
\frac{\partial m}{\partial\xi}=-\frac{2m}{d},
\qquad
\frac{\dd K}{\dd m}=\frac{E-(1-m)K}{2m(1-m)},
\]
where formulas containing \(1/m\) are understood by continuity at \(m=0\).
Since \(M_2=-\partial_\xi M_1\),
\[
M_2=\frac{2E(m)}{\pi d^2(1-m)}
=\frac{2E(m)}{\pi\xi(\xi+8\chi)}.
\]
The condition \(M_2=1\) and \(G=M_1\) give Eq.~\eqref{eq:KE}. At \(\chi=0\),
\(\xi=G=1\) and \(\Phi(0,\tau)=1+\tau\).

\paragraph{Zero-redistribution threshold.}

Combining the expressions for $G$ and $M_2=1$
gives
\[
\frac{\xi}{G}
=\frac{E(m)}{(1+\sqrt m)K(m)}.
\]

As $\chi$ increases, $m$ increases from
$0$ to $1$. Since $E(m)$ decreases and
$K(m)$ increases, $\xi/G$ decreases
strictly from $1$ to $0$.

Thus, for each $-1<\tau<0$, the equation
$\Phi=\xi+\tau G=0$ has a unique solution
$\chi_0(\tau)$, with $\Phi<0$ for
$\chi>\chi_0(\tau)$.

\paragraph{Strong diffusion.}
As \(\chi\to\infty\), \(M_2\) is dominated by the neighborhood of \(k=0\),
where \(\omega(k)=|k|^2+O(|k|^4)\). The continuum contribution is
\[
\int_{\mathbb R^2}\frac{\dd^2k}{(2\pi)^2}
\frac1{(\xi+\chi|k|^2)^2}
=\frac1{4\pi\chi\xi},
\]
so \(\xi\sim(4\pi\chi)^{-1}\). The constant in \(G\) requires the full
lattice Green function. From
\[
1-m=\frac{\xi(\xi+8\chi)}{(\xi+4\chi)^2}\sim\frac{1}{8\pi\chi^2},
\qquad
K(m)=\log\frac4{\sqrt{1-m}}+O\bigl((1-m)\log(1-m)\bigr),
\]
and \(E(m)=1+O\bigl((1-m)\log(1-m)\bigr)\), one obtains
\[
\xi=\frac{1}{4\pi\chi}+O\!\left(\frac{\log\chi}{\chi^3}\right),
\]
\[
G=\frac{\log\chi+C_0}{2\pi\chi}+O\!\left(\frac{\log\chi}{\chi^3}\right),
\qquad C_0=\log(8\sqrt{2\pi}),
\]
which is Eq.~\eqref{eq:strong}.

\section{Finite-size criterion and weak-anticorrelation asymptotics}\label{sm:finite-size}

\paragraph{Finite-size criterion.}
The smallest nonzero square-lattice frequency is
\[
\omega_{\min}(L)=4\sin^2\left(\frac{\pi}{L}\right)
\simeq\frac{4\pi^2}{L^2}.
\]
The long-wavelength contribution to the endpoint equation is resolved when
\[
\frac{\chi\omega_{\min}(L)}{\xi(\chi)}\ll1.
\]
Using \(\xi(\chi)\simeq(4\pi\chi)^{-1}\) gives \(L\gg4\pi^{3/2}\chi\).
At fixed \(L\), every nonuniform Laplacian eigenvalue is at most
\(-\omega_{\min}(L)<0\), so sufficiently strong diffusion drives the real
parts of the spectrum on \(\ee^\perp\) to \(-\infty\).

\paragraph{Weak anticorrelation.}
Set \(t=-\tau>0\). From Eq.~\eqref{eq:strong},
\[
4\pi\chi\,\Phi(\chi,-t)=1-2t(\log\chi+C_0)
+O\!\left(\frac{\log\chi}{\chi^2}\right).
\]
Expanding the exact elliptic-integral formulas and differentiating gives
\[
\xi'=-\frac{1}{4\pi\chi^2}
+O\left(\frac{\log\chi}{\chi^4}\right),
\]
\[
G'=\frac{1-\log\chi-C_0}{2\pi\chi^2}
+O\left(\frac{\log\chi}{\chi^4}\right),
\]
hence
\[
\begin{aligned}
2\pi\chi^2\partial_\chi\Phi(\chi,-t)
={}&-\frac12+t(\log\chi+C_0-1)\\
&+O\left(\frac{(1+t)\log\chi}{\chi^2}\right).
\end{aligned}
\]

As $t\to0^+$, both $\chi_0$ and $\chi_\star$
diverge. Indeed, for any fixed $A>0$,
the monotonicity of $\xi$ and $G(\chi)\leq1$
give
\[
\Phi(\chi,-t)\geq\xi(A)-t,
\qquad 0\leq\chi\leq A.
\]
Thus, for sufficiently small $t$, the edge
remains positive over any fixed finite range
of $\chi$, whereas it becomes negative at
large diffusion. Both the zero crossing and
the global minimum therefore move toward
arbitrarily large diffusion strengths.

\emph{Sign change.} In the large-\(\chi\) region, the derivative of
\(4\pi\chi\Phi\) with respect to \(\log\chi\) is
\(-2t+O(\log\chi/\chi^2)<0\). There is therefore a single zero
\(\chi_0\) for small \(t\), and it satisfies
\(\log\chi_0+C_0=1/(2t)+o(1)\):
\[
\chi_0\sim\frac{e^{1/(2t)}}{8\sqrt{2\pi}}.
\]

\emph{Minimum.} At an interior minimizer, the stationary equation first
gives \(t\log\chi_\star\to1/2\). The remainder is then \(o(t)\), and the
same equation gives \(\log\chi_\star+C_0-1=1/(2t)+o(1)\):
\[
\chi_\star\sim\frac{e}{8\sqrt{2\pi}}e^{1/(2t)}\sim e\,\chi_0.
\]
This control of the differentiated remainder fixes the multiplicative
prefactor. Substituting into the expansion of \(\Phi\),
\[
4\pi\chi_\star\Phi(\chi_\star,-t)=1-2t\left(\frac{1}{2t}+1\right)
+o(t)=-2t+o(t),
\]
so that
\[
\Phi(\chi_\star,-t)\sim-\frac{t}{2\pi\chi_\star}.
\]
These are the results quoted in the Letter.

\section{One-dimensional cycle}\label{sm:cycle}

For the cycle Laplacian, \(\omega(k)=2-2\cos k\), averaged over
\([-\pi,\pi]\). Define
\[
D=\xi(\xi+4\chi),\qquad d=\xi+2\chi.
\]
At \(v=0\), \(\langle(s+\chi\omega)^{-2}\rangle\) diverges for
\(-4\chi\leq s\leq0\) (as \(s^{-3/2}\) when \(s\downarrow0\)) and decreases
strictly to zero for \(s>0\). The argument of Sec.~\ref{sm:boundary} therefore again
places the right edge at \(v=0\). The right-edge equations reduce to
\[
G=D^{-1/2},\qquad M_2=dD^{-3/2}=1.
\]
Consequently,
\[
\xi\sim2^{-4/3}\chi^{-1/3},\qquad
G\sim2^{-1/3}\chi^{-1/3},\qquad
\frac{\xi}{G}\to\frac12,
\]
and, for \(\tau\ne-1/2\),
\[
\Phi\sim2^{-4/3}(1+2\tau)\chi^{-1/3}.
\]
The edge thus approaches zero from below for \(\tau<-1/2\). At
\(\tau=-1/2\), using \(G=D/d\),
\[
\Phi=\xi-\frac G2
=\frac{\xi^2}{2(\xi+2\chi)}>0,
\]
so it approaches zero from above.

\end{document}